\documentclass[english,twocolumn,aps,prl,showpacs]{revtex4-1}

\usepackage[T1]{fontenc}
\usepackage[latin1]{inputenc}
\usepackage{graphicx}
\usepackage{amssymb}
\usepackage{epsfig}

\makeatletter

\input{epsf}

\makeatother

\usepackage{babel}
\makeatother

\begin{document}

\title{Universal behavior of pair correlations in a strongly interacting Fermi gas}

\author{E. D. Kuhnle, H. Hu, X.-J. Liu, P. Dyke, M. Mark, P. D. Drummond, P. Hannaford, and C. J. Vale}
\affiliation{\ ARC Centre of Excellence for Quantum-Atom Optics, Centre for Atom Optics and Ultrafast Spectroscopy, \\
Swinburne University of Technology, Melbourne 3122, Australia}

\date{\today}

\begin{abstract}
We show that short-range pair correlations in a strongly interacting Fermi gas follow a simple universal law described by Tan's relations.  This is achieved through measurements of the static structure factor which displays a universal scaling proportional to the ratio of Tan's contact to the momentum ${\cal C}/q$.  Bragg spectroscopy of ultracold $^6$Li atoms from a periodic optical potential is used to measure the structure factor for a wide range of momenta and interaction strengths, providing broad confirmation of this universal law.  We calibrate our Bragg spectra using the $f$-sum rule, which is found to improve the accuracy of the structure factor measurement.
\end{abstract}


\maketitle

Universality is a remarkable property of strongly interacting systems of fermions \cite{heiselberg,ho,hunatphys}.  Universality means that all dilute Fermi systems with sufficiently strong interactions behave identically on a scale given by the average particle separation.  With the discovery of universality in the Bose-Einstein condensate (BEC) to Bardeen-Cooper-Schrieffer (BCS) crossover, ultracold Fermi gases near Feshbach resonances have become a central topic in atomic physics \cite{jochim,greiner,zwierlein,kinast,partridge,nascimbene,navon}.  Strongly interacting Fermi systems arise in a wide variety of settings, from astrophysical to nuclear and condensed matter systems.  One can therefore study universality in ultracold atomic gases to help understand other strongly interacting Fermi superfluids, taking advantage of the ability to precisely control the atom-atom interactions.

Understanding these strongly interacting Fermi gases, however, poses significant challenges \cite{giorgini}.  In 2005 Shina Tan made dramatic progress by deriving several exact relations for Fermi gases in the BEC-BCS crossover, which relate the microscopic properties to bulk thermodynamic quantities \cite{tan1,tan2,tan3}.  These exact relations are applicable in broad circumstances: zero or finite temperatures, superfluid or normal phases, homogeneous or trapped systems, and in few or many-body systems.  

In this letter, we experimentally verify a universal relation for short-range pair correlations \cite{tan1} using Bragg spectroscopy.  This is achieved through measurements of the static structure factor, given by the Fourier transform of the pair correlation function \cite{combescot}.  The structure factor of a unitary Fermi gas has an exact scaling with the ratio of Tan's contact to the momentum ${\cal C}/q$.  For systems with finite scattering length, we also confirm the first order correction to the universal law.  Our measurements are compared to new calculations for the contact based on a recently developed below-threshold Gaussian pair fluctuation theory \cite{hueur}.

The contact $\cal{C}$ in a two component Fermi gas quantifies the likelihood of finding two fermions with opposite spin close enough to interact with each other.   In systems where the range of the interaction potential is negligible, this single parameter encapsulates all of the information required to determine the many-body properties \cite{braaten,zhang}.   $\cal{C}$ depends on the $s$-wave scattering length, density and temperature of the system.  Tan showed that the internal energy of a gas across the BEC-BCS crossover can be expressed as a functional of the momentum distribution which has a ${\cal C}/q^4$ dependence at large momentum $q$ and that the pair correlation function diverges as ${\cal C}/r^2$ at short distance $r < 1/k_F$, where $k_F$ is the Fermi wavevector \cite{tan1,combescot}.  Tan also derived the adiabatic relation $dE/d(-1/a)=\hbar ^2{\cal C}/(4\pi m)$, where $m$ is the atomic mass, giving the change in the total energy $E$ due to an adiabatic change in the scattering length \cite{tan2} and extended the virial theorem to finite $a$ and imbalanced mixtures \cite{tan3}.  The contact $\cal{C}$ was first extracted \cite{werner} from the number of closed-channel molecules determined through photo-association \cite{partridge} and the adiabatic and virial Tan relations were very recently verified experimentally \cite{stewartgaebler}.  We will generally refer to the dimensionless contact ${\cal I}$ given by ${\cal C}/(N k_F)$ where $N$ is the number of particles.

Short-range structure in a quantum fluid depends upon the relative wave-function of the interacting particles, in this case fermions in different spin states.  In a two-component (spin-up/spin-down) Fermi gas with contact interactions this is given by $\psi_{\uparrow \downarrow}(r) \propto 1/r - 1/a$, where $a$ is the $s$-wave scattering length.  Starting from this wavefunction, Tan showed that the spin-antiparallel pair correlation function is given by Eq.$\:$(1) which includes the contact as a pre-factor \cite{tan1}
\begin{equation}
g_{\uparrow \downarrow}^{(2)}(r) \rightarrow \frac{{\cal I}}{16\pi ^2} \left(\frac{1}{r^2}-\frac{2}{ar}\right).
\end{equation}

Pair correlation functions are difficult to measure directly in ultracold gases; however, it is possible to measure macroscopic quantities which depend on correlation functions in a well defined way.  A prime example is the static structure factor, $S(k)$, which is given by the Fourier transform of $g^{(2)}(r)$ ($q = \hbar k$ is the probe momentum).  In a two component Fermi gas with an equal number of particles in each state $N$, the structure factor consists of two components, corresponding to correlations between particles in the same state and particles in different states $S(k) = S_{\uparrow \uparrow}(k) + S_{\uparrow \downarrow}(k)$.  When the momentum is much larger than the Fermi momentum, particles in the same state will be uncorrelated and $S_{\uparrow \uparrow}(k \gg k_F) \simeq 1$ so all variation in $S(k)$ will then be due to changes in $S_{\uparrow \downarrow}(k)$ \cite{combescot,veeravalli}.  The Fourier transform of Eq.$\:$1 yields the following for $S_{\uparrow \downarrow}(k)$
\begin{equation}
S_{\uparrow \downarrow} \left( k \gg k_F\right) = \frac{\cal I}{4} \frac{k_F}{k}\left[1-\frac{4}{\pi k_F a} \left(\frac{k_F}{k} \right) \right],  \label{TanSSF}
\end{equation}
which has a straightforward dependence upon ${\cal I}$, the relative probe momentum $k/k_F$, and the interaction strength $1/(k_F a)$.  At unitarity, $a \rightarrow \infty$, the second term vanishes and $S(k)$ varies linearly with $k_F/k$.  Rewriting Tan's relation in this way points to a method to experimentally verify Tan's universal relation for pair correlations between spin-up/spin-down fermions since the static structure factor can readily be measured using Bragg spectroscopy.

\begin{figure}[htp]

\begin{centering}
\includegraphics[clip,width=0.45\textwidth]{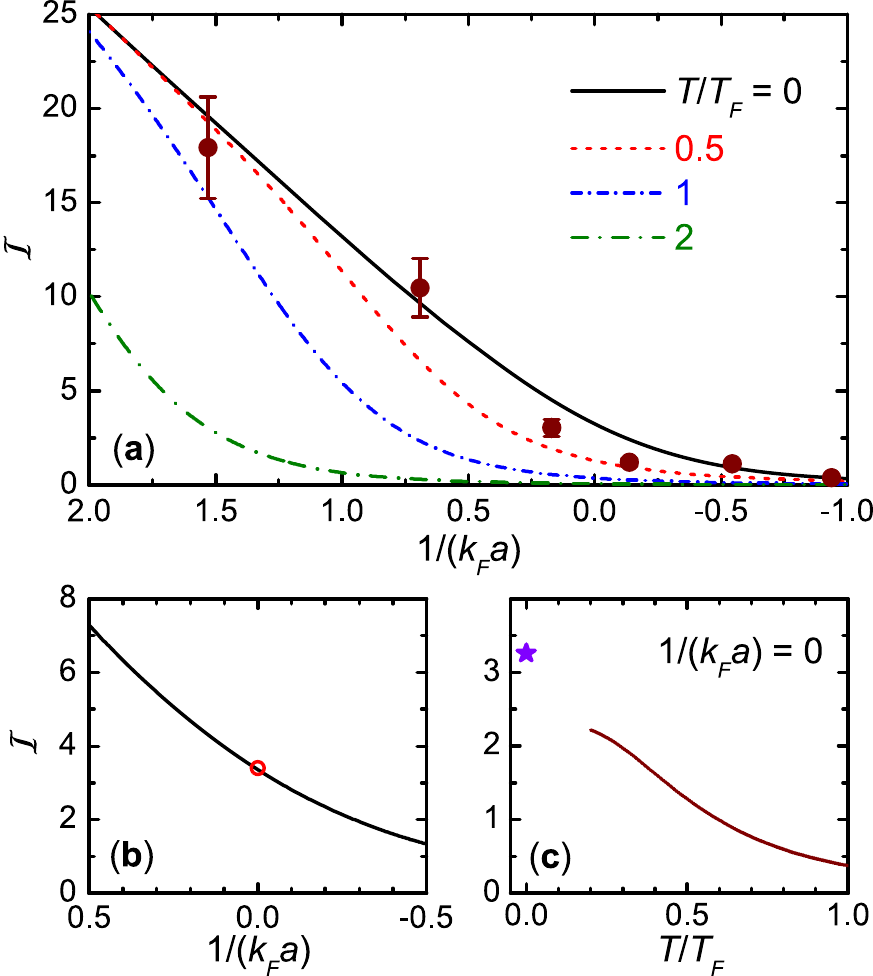}
\par\end{centering}

\caption{Theoretical prediction for Tan's contact in the BEC-BCS crossover. (a) Evolution of contact for a trapped Fermi gas with $1/(k_Fa)$ at different temperatures.  Points are experimental data from refs \cite{partridge,werner}.  The $T$ = 0 curve is calculated using the contact for a homogeneous Fermi gas (shown in b) and a local density approximation.  Finite temperature contact is determined from a cluster expansion theory \cite{liu}. (b) Zero temperature contact in free space, obtained from a Gaussian pair fluctuation theory \cite{hueur}. Red circle shows the contact from QMC for pair correlation functions at unitarity \cite{combescot}. (c) Temperature dependence of the trapped contact at unitarity.  With decreasing $T$, contact increases rapidly towards the $T$ = 0 result (purple star).}

\label{fig1}
\end{figure}

Apart from the contact, all parameters in Eq.$\:$(2) are easily determined or set by experimental parameters.  We employ new methods to calculate the contact which overcome the need for interpolation schemes between the limiting BEC and BCS cases used in earlier studies \cite{werner}.  To find the low temperature contact we first calculate the ground state energy across the BEC-BCS crossover using a recently developed below-threshold Gaussian pair fluctuation theory \cite{hueur}, which has shown excellent agreement with thermodynamic measurements \cite{nascimbene,navon,luo} and quantum Monte Carlo (QMC) simulations \cite{astrakharchik}.  We then use Tan's adiabatic theorem to find the contact.  The high temperature $(T \gtrsim 0.5 \, T_F)$ contact is calculated from a quantum cluster expansion \cite{liu}.  Figure \ref{fig1} shows the calculated contact ${\cal I}$ for a range of temperatures through the BEC-BCS crossover, along with the results from photo-association (Fig.$\:$\ref{fig1}a) \cite{partridge,werner}.  The contact increases monotonically from the BCS to BEC regimes and is highest at low temperatures.  Combining these calculated results for the contact with Eq.$\:$(2) for high momentum transfer Bragg scattering, we obtain a direct, quantitative prediction that is readily tested experimentally.  

Inelastic scattering experiments are a well established technique to probe both the dynamic and static structure factors of many-body quantum systems \cite{pines}.  Ultracold atoms are highly amenable to inelastic scattering through Bragg spectrosocpy which has previously been used to measure both the dynamic \cite{stenger,zambelli} and static structure factors \cite{stamper,steinhauer} of atomic Bose-Einstein condensates.  In ultracold Fermi gases Bragg spectroscopy was used to measure both the dynamic and static structure factors over the BEC-BCS crossover, albeit at a single momentum \cite{veeravalli}.  The ability to vary the momentum differentiates Bragg spectroscopy from rf spectroscopy \cite{schunck,stewart}, providing access to a broad range of the excitation spectrum and avoiding final state interaction effects as no third atomic state is involved.  

To demonstrate the validity of Eq.$\:$(2) we need to measure $S(k)$ for a range of momenta ($k/k_F$) and at different values of the interaction parameter $1/(k_F a)$.  The starting point of our experiments are clouds containing $N \approx 3 \times 10^5$ $^6$Li atoms in an equal mixture of the two lowest lying ground states $| F = 1/2, m_F = \pm 1/2 \rangle$ evaporatively cooled in a single beam optical dipole trap ($\lambda = 1075$ nm) at a magnetic field of 834 G \cite{fuchs}.  Next, we adiabatically ramp up a second far detuned laser ($\lambda = 1064$ nm) which intersects the first trapping beam at an angle of 74$^\circ$ forming a crossed beam dipole trap.  By appropriately selecting the intensities of each of the two trap lasers we tune the mean harmonic confinement frequency of the crossed trap $\bar{\omega}$ over the range $\bar{\omega} = 2 \pi \times (38 \rightarrow 252)$ s$^{-1}$ while the aspect ratio, $\omega_{x,y}/\omega_z$, varies from 3.4 to 16.  Controlling the trap frequencies in this way allows us to tune the atom density and hence the Fermi wavevector over a broad range.  Once in the final crossed beam trap, we ramp the magnetic field to select the scattering length $a$ which gives the desired value of $1/(k_Fa)$.  This allows us to tune $k_F$ and $1/(k_Fa)$ independently.  The cloud is then held at the final magnetic field for a time $\tau \gg 10 \, \bar{\omega}^{-1}$ to reach equilibrium before applying the Bragg pulse and imaging at that magnetic field.

Bragg scattering is achieved by illuminating these  $^6$Li clouds with two counter propagating laser beams with a very small frequency difference $\delta$.  This creates a standing wave which moves with velocity $\delta/k_{Br}$ where $k_{Br}=2 \pi / \lambda_{Br}$ is the wave-vector and $\lambda_{Br} = 671$ nm is the wavelength of the Bragg lasers.  When the energy difference between the two Bragg lasers equals the kinetic energy associated with a two-photon recoil, resonant Bragg scattering occurs (i.e.$\:\delta = 2 \hbar k_{Br}^2 / m$).  The Bragg lasers are approximately 2 GHz red-detuned from the two ground hyperfine states which is large compared to the $\sim$ 80 MHz splitting between the $|1/2,\pm1/2\rangle$ states.  This means that both states are coupled approximately equally to the Bragg beams (to within 4$\%$).

To measure the static structure factor $S(k)$ we begin by recording a Bragg spectrum.  This consists of measuring the momentum transferred to the cloud by the Bragg lasers as a function of $\delta$.  The transferred momentum is directly proportional to the resultant center of mass displacement $\Delta X(k,\delta)$ measured after 2 ms time of flight \cite{veeravalli}. A Bragg spectrum for a unitary gas is shown in the inset of Fig. 2 for $k/k_F = 8.5$ as a function of the Bragg frequency $\delta$.  At this momentum, the pair and free-atom excitations are clearly distinguished at frequencies of $\sim 150$ kHz and $\sim 300$ kHz, respectively.  The Bragg pulse duration, 50 $\mu$s, is short compared to the two-photon Rabi cycling period, ensuring that spectra are obtained in the linear response regime.
\begin{figure}[htp]

\begin{centering}
\includegraphics[clip,width=0.45\textwidth]{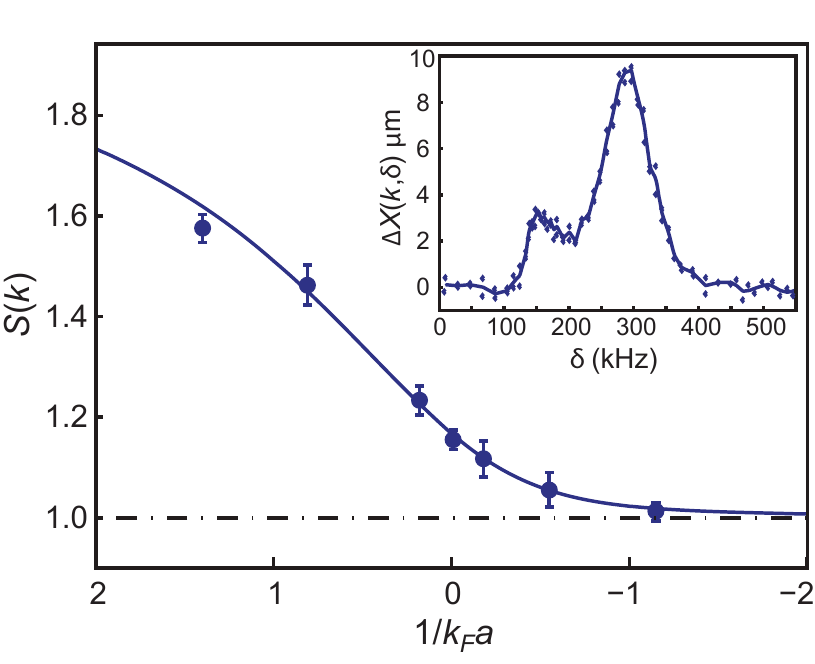}
\par\end{centering}

\caption{Static structure factor in the BEC-BCS crossover.  Experimental points were obtained by integrating Bragg spectra normalised via the $f$-sum rule for $k/k_F$ = 4.8.  Error bars are due to shot to shot atom number fluctuations and uncertainties in measuring center of mass.  The theoretical line is a zero temperature result, calculated by interpolating the near-resonance structure factor Eq. (2) with the asymptotic results in the BCS and BEC limits.  Inset: A Bragg spectrum obtained at $1/(k_Fa)$ = 0 and $k/k_F$ = 8.5 showing center of mass displacement $\Delta X(k,\delta)$ versus Bragg frequency $\delta$.  Points are single shots and the line is a guide to the eye.  The pair ($\delta \sim$ 150 kHz) and free atom peaks ($\delta \sim$ 300 kHz) are clearly distinguished.}

\label{fig2}
\end{figure}

The center of mass displacement, $\Delta X(k,\delta)$, is proportional to the convolution of the spectral content of the Bragg pulse and the dynamic structure factor of the gas $S(k,\delta)$ \cite{brunello}.  The proportionality constant depends upon the two-photon Rabi frequency $\Omega_{Br}$ which can be difficult to accurately measure.  Integrating $\Delta X(q,\delta)$ over $\delta$ yields a number proportional to the static structure factor, $S(k) = \hbar N^{-1} \int S(k,\delta) d\delta$, but $\Omega_{Br}$ remains unknown.  We overcome the need to find $\Omega_{Br}$ by invoking the $f$-sum rule \cite{pines}: $NE_r = \int S(k,\delta) \delta d \delta$, where $E_r = 2 \hbar^2 k_{Br}^2/m$ is the two-photon recoil energy.  As both $S(k)$ and the $f$-sum rule involve $N$, we normalise the area under each measured spectra using
\begin{equation}
S(k) = \frac{2 \hbar k_{Br}^2}{m} \frac{\int \Delta X(k,\delta) d\delta}{\int \Delta X(k,\delta) \delta d\delta},
\label{fsum}
\end{equation}
where the constant involving $\Omega_{Br}$ appears before each integral and therefore cancels.  Equation (3) is absolute, requiring only knowledge of the recoil energy, which can be determined with high accuracy, leading to an accurate measure of $S(k)$.  In Fig.$\:$2 we plot $S(k)$ obtained in this way for $k/k_F = 4.8$ which shows excellent agreement with the zero temperature theoretical calculation.
\begin{figure}[htp]

\begin{centering}
\includegraphics[clip,width=0.45\textwidth]{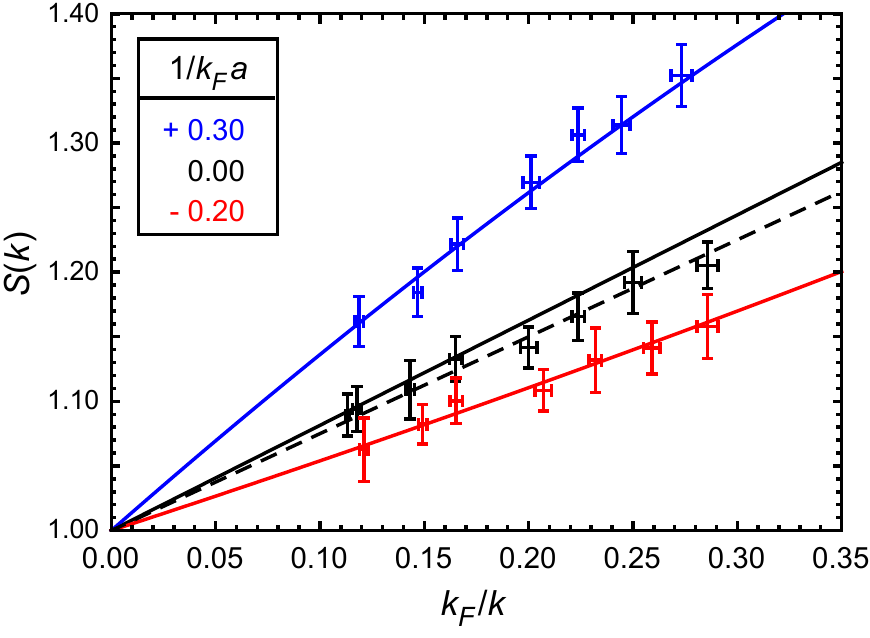}
\par\end{centering}

\caption{Universal dependence of the static structure factor of a strongly interacting Fermi superfluid.  Measured and calculated static structure factor versus $k_F/k$ for $1/(k_Fa)$ = +0.3, 0.0, and -0.2.  Bragg momentum $k$ is fixed while $k_F$ is varied by changing the mean trapping frequency $\bar{\omega}$.  Vertical error bars are due to atom number fluctuations and uncertainties in measuring the center of mass and horizontal error bars are due to atom number fluctuations and uncertainties in $\bar{\omega}$.  Solid lines are the zero temperature theory and the dashed line is a straight line fit to the $1/(k_Fa)$ = 0 data yielding a slope of $0.75 \pm 0.03$.}

\label{fig3}
\end{figure}

Next we come to the measurement of $S(k)$ as a function of $k/k_F$.  At unitarity we expect $S(k)$ to vary linearly with $k_F/k$ according to Eq.$\:$(2).  Rather than changing $k$ directly by varying the angle between the two Bragg beams, we change $k_F = (48N)^{1/6} \sqrt{m \bar{\omega}/\hbar}$ by varying $\bar{\omega} = (\omega_x \omega_y \omega_z)^{1/3}$, the geometric mean frequency of the optical dipole trap.  Universality allows us to change the relative length scale being probed simply by changing the density of the gas.  Using the crossed beam optical trap described above we can tune $\bar{\omega}$ such that the Fermi wavevector can be tuned anywhere over the range $k_F$ = 2.1 $\mu$m$^{-1}$ $\rightarrow$ 5.3 $\mu$m$^{-1}$, or $k/k_F$  = 3.5 $\rightarrow$ 9.1.  A sequence of Bragg spectra were obtained for gases at three values of $1/(k_Fa)$ = +0.3, 0.0 and -0.2, while varying $k_F$.  From these, $S(k)$ was extracted and the results are plotted in Fig.$\:$3.  The solid lines are the prediction from Eq.$\:$(2) using the zero temperature contact with no free parameters.  The experimental points closely follow the theory, confirming the exact analytic dependence of $S_{\uparrow \downarrow}(k) \cong S(k) - 1$ on $q$, Eq.$\:$(2).  At unitarity $1/(k_Fa)$ = 0, the dependence on $a$ vanishes and a straight line fit (dashed line) shows the simple universal behavior of fermionic pairing.  The fitted slope of $0.75 \pm 0.03$ is slightly below the zero temperature prediction of 0.81 which may be due to reduced pairing at the finite temperature ($T/T_F = 0.10 \pm 0.02$ at unitarity).  At $1/(k_Fa)$ = -0.2 the temperature will be lower following the adiabatic magnetic field sweep \cite{carr} while at $1/(k_Fa)$ = +0.3 the temperature will be higher but pairing takes place at much higher temperatures.  At $T \ll T_{C} \: (\simeq 0.2 \, T_{F})$, phonons dominate the excitations and the contact should increase as $(T/T_{F})^{4}$ \cite{yu}.  In the relevant temperature window, we estimate this increase to be only $0.1\%$ which could easily be negated by normal single-particle excitations localized at the cloud edge.  At $1/(k_Fa)$ = +0.3 the data depart from a straight line displaying the downward curvature consistent with the first order term in Eq.$\:$(2).  A similar upward curvature is seen at $1/(k_Fa)$ = -0.2.  Our simple relation Eq.$\:$(2) is seen to accurately describe $S(k)$ on both sides of the Feshbach resonance demonstrating the wide applicability of the Tan relations.

In summary, we have shown that the structure factor of a strongly interacting ultracold Fermi gas follows a universal law which is a direct consequence of Tan's relation for the pair correlation functions.  Our measurements provide one of the first demonstrations of a broadly applicable exact result for Fermi gases in the BEC-BCS crossover.  This work opens the way to a complete temperature and interaction dependent map of the contact through the BEC-BCS crossover and may provide a new means for obtaining the equation of state.

This work is supported by the Australian Research Council Centre of Excellence for Quantum-Atom Optics and Discovery Projects DP0984522 and DP0984637.

\end{document}